\documentclass[twocolumn,showpacs,preprintnumbers,amsmath,amssymb,aps,prb]{revtex4}

\usepackage{graphicx}
\usepackage{dcolumn}
\usepackage{bm}
\usepackage{subfigure}
\usepackage{natbib}
\usepackage{multirow}
\usepackage{xcolor}

\begin{document}

\title{Spin injection efficiency at metallic interfaces probed by THz emission spectroscopy}

\author{Jacques Hawecker$^{1}$, T. H. Dang$^{2}$, Enzo Rongione$^{2}$, James Boust$^{2}$, Sophie Collin$^{2}$, Jean-Marie George$^{2}$, Henri-Jean Drouhin$^{3}$, Yannis Laplace$^{3}$, Romain Grasset$^{3}$, Jingwei Dong$^{3}$, Juliette Mangeney$^{1}$, Jerome Tignon$^{1}$, Henri Jaffr\`es$^{2}$, Luca Perfetti$^{3}$\footnote{The authors to whom the correspondence show be addressed are luca.perfetti@polytechnique.edu, henri.jaffres@cnrs-thales.fr and sukhdeep.dhillon@phys.ens.fr} and Sukhdeep Dhillon$^{1}$}
\affiliation
{$^{1}$ Laboratoire de Physique de l'Ecole normale sup\'rieure, ENS, Universit\'e PSL, CNRS, Sorbonne Universit\'e, Universit\'e de Paris, F-75005 Paris, France}
\affiliation
{$^{2}$ Unit\'e Mixte de Physique, CNRS, Thales, Universit\'e Paris-Sud, Universit\'e Paris-Saclay, F-91767 Palaiseau, France}
\affiliation
{$^{3}$ Laboratoire des Solides Irradi\'{e}s, CEA/DRF/lRAMIS, Ecole Polytechnique, CNRS, Institut Polytechnique de Paris, F-91128 Palaiseau, France}

\begin{abstract}

Terahertz (THz) spin-to-charge conversion has become an increasingly important process for THz pulse generation and as a tool to probe ultrafast spin interactions at magnetic interfaces. However, its relation to traditional, steady state, ferromagnetic resonance techniques is poorly understood. Here we investigate nanometric trilayers of Co/X/Pt (X=Ti, Au or Au0:85W0:15) as a function of the 'X' layer thickness, where THz emission generated by the inverse spin Hall effect is compared to the Gilbert damping of the ferromagnetic resonance. Through the insertion of the 'X' layer we show that the ultrafast spin current injected in the non-magnetic layer defines a direct spin conductance, whereas the Gilbert damping leads to an effective spin mixing-conductance of the trilayer. Importantly, we show that these two parameters are connected to each other and that spin-memory losses can be modeled via an effective Hamiltonian with Rashba fields. This work highlights that magneto-circuits concepts can be successfully extended to ultrafast spintronic devices, as well as enhancing the understanding of spin-to-charge conversion processes through the complementarity between ultrafast THz spectroscopy and steady state techniques.

\end{abstract}


\maketitle

\section{Introduction}

When a pure spin current pass through materials with large spin-orbit coupling, it can generate a transverse charge current \cite{Saitoh,Valenzuela} by means of the Inverse Spin-Hall-Effect (ISHE). A flurry of activity on this topic has been motivated by the intimate relation between ISHE and the direct SHE \cite{Wunderlich}. The latter can be very efficiently employed to generate a spin transfer torque capable of switching the magnetization of ferromagnetic thin films~\cite{Gambardella,Liu}. Most experiments in this field have been performed by spin-pumping \textit{via} ferromagnetic resonance while some works have investigated the ultrafast regime \cite{Garello,Decker}. More recently, other authors have proven that ISHE can be employed to generate an intense THz radiation. This breakthrough highlighted that interfaces leading to large spin transfer torque are also excellent emitters of electromagnetic waves~\cite{Seifert1,Yang,Wu,Nenno,Dang2020}. Theoretical simulations based on superdiffusive transport equations have successfully reproduced the observed emission \cite{Schneider,Battiato}. However, these frameworks do not cover the impact of the electronic transmission at interfaces, neither the discussion of the particular role of the interfacial spin-orbit fields originating from charge transfer and symmetry breaking~\cite{Buhrman2}. To this end, a tighter connection with steady state spintronics is highly desirable. For example, the magnetocircuits analogies are widely employed to define the efficiency of the spin-to-charge conversion \cite{Halperin,Brataas}. An extension of such formalism to impulsive excitations has been discussed in the case of ultrafast spin-Seebeck effect \cite{Seifert3} but not yet for the spin current injected from ferromagnetic transition metals.

\vspace{0.1in}

In the field of spintronics, the optimal efficiency of spin orbit torque (SOT) requires the engineering of metallic interfaces favoring a higher spin-current generation. Recently, many authors have tackled this issue by inserting transion metals \cite{Avci,Park} or noble metals \cite{Beach} between cobalt (or CoFe, CoFeB) and platinum. Experiments with different interlayers have shown clear correlations between the spin-transfer-torque, magnetoresistance \cite{Avci}, perpendicular anisotropy \cite{Avci,Buhrman2} and spin memory loss \cite{Kelly2}. Here, we investigate this topic by comparing the THz emission efficiency detected by Time Domain Spectroscopy (TDS) with Ferro-Magnetic Resonance (FMR) spectroscopy of trilayers Co/X/Pt. The insertion of an atomically thick  interlayer of X=Ti, Au or Au$_{0.85}$W$_{0.15}$ modifies the capability of the interface to generate spin currents. This property affects, on the same footing, the emission of electromagnetic radiation as well as the Gilbert damping of the multilayer. Our result show that the THz-TDS emission spectroscopy is contactless and non-destructive method that can give an accurate and reliable estimate of the spin-injection efficiency at spintronic interfaces. We discuss the data in the magnetocircuit formalism \cite{Halperin,Brataas,Seifert3}, by assuming that spin-currents follow the evolution of magnetic fluctuations. The average spin-conductance $(g^{\uparrow}+g^{\downarrow})/2$ characterizes the ultrafast currents in platinum \cite{Seifert3} whereas an effective spin conductance $g^{\uparrow \downarrow}_{\textrm{eff}}$ describes the damping torque in the ferromagnetic material ( in our case is cobalt )\cite{Halperin}. These two quantities would be proportional to each other if the spin flow was conserved at the interface \cite{Brataas}. In reality, strong spin-orbit assisted scattering processes generate a sink of angular momentum and limit the spin flow that can propagate in platinum \cite{Jaffres,Silva,Kelly1,Buhrman1}. The comparison between $(g^{\uparrow}+g^{\downarrow})/2$ and $g^{\uparrow \downarrow}_{\textrm{eff}}$ shows that the fraction of spin current lost at the interface \cite{Kelly1} is proportional to the spin conductance and may exceed 40\% in the Co/Pt bilayer. Our measurements highlight that passivation of the interface by different compounds follows a common trend and suggests the existence of a general relation between spin memory loss and spin-conductance. We prove this claim by choosing inter-layer materials with very different properties: Ti is more chemically reactive and has small spin-orbit coupling whereas Au and Au$_{0.85}$W$_{0.15}$ are less chemically reactive and hosts a larger spin-orbit interaction (especially the Au:W alloy).

\begin{figure}[h]
\includegraphics[width=0.9\columnwidth]{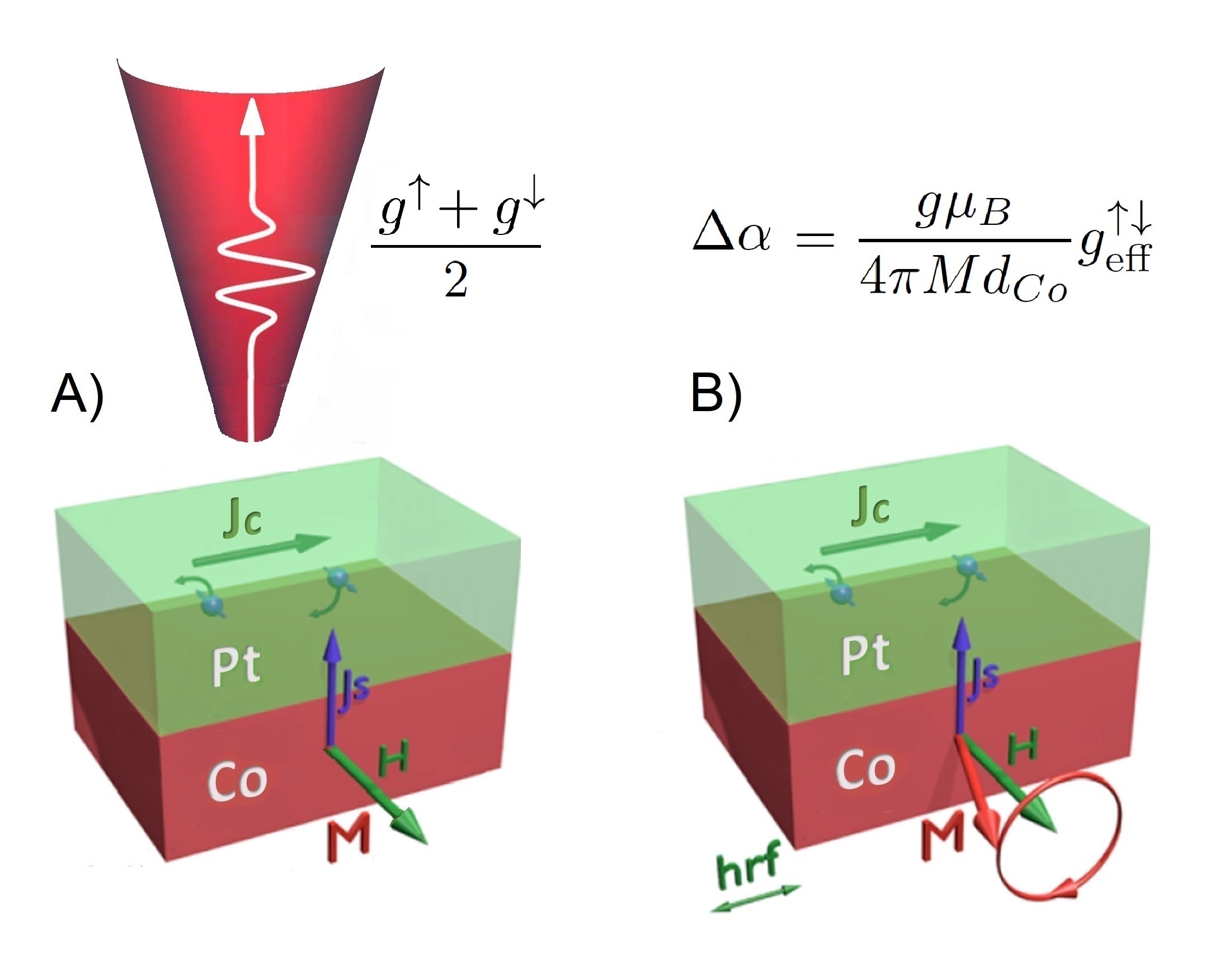}
\caption{A) Detection of spin-to-charge conversion in a spintronic emitter. The cobalt layer has magnetization $\vec M$ parallel to the external magnetic field $\vec H$ and is in contact with the platinum layer. An ultrashort laser pulse photoexcites the sample generates, in the Pt side of the interface, a spin current $ \mathcal{\vec J}_s$ that is proportional to the direct spin conductance $(g^{\uparrow}+g^{\downarrow})/2$. The inverse spin Hall effect of Pt leads to a transverse charge current $\mathcal{\vec J}_c$. Being shorter than one picosecond, the $\mathcal{\vec J}_c$ pulse emits radiation in the THz spectral range. B) Effect of the platinum layer on the ferromagnetic resonance of the underlying cobalt. The magnetization precession driven by a radiofrequency field $\vec h_{rf}$ induces a spin current. The increase of Gilbert damping due to the $\mathcal{\vec J}_s$ injection in the Pt layer is proportional to the effective spin-mixing conductance.}
\label{Sketch}
\end{figure}

\section{General Framework of spintronic THz emission}

A framework building on few hypothesis connects the spin conductance to the emitted THz radiation. In the thin film limit, the THz electric field of a plane wave at the surface of the sample is given by $\vec  E_{T}(\omega)=eZ \int \mathcal{\vec J}_c(\omega,z)dz$. This expression links $\vec  E_{T}(\omega)$ to the charge current density $\mathcal{\vec J}_c$ via an effective impedance \cite{Seifert1,Seifert2,Papaioannou}:
\begin{equation}
Z=\frac{Z_0}{1+n+Z_0\int \sigma (z)dz},
\label{Imp}
\end{equation}
where $n$ is the refractive index of the substrate, $z$ is the coordinate perpendicular to the interface, $Z_0=377 \Omega$ is the vacuum impedance and $\int\sigma(z)dz$ is the local conductivity integrated over the total thickness of the multilayer. The charge current $\mathcal{\vec J}_c(z)$ arises in platinum because of the inverse-spin-Hall-effect acting on the spin current flow $\mathcal{\vec J}_s(z)$ along the normal direction to the film plane. The latter decreases exponentially over a distance equal to the spin diffusion length. It follows that:
\begin{equation}
\int_{0_+}^{d_{Pt}} \mathcal{\vec J}_c(z)dz=\mathcal{J}_s(0_+)(\vec e_n  \times \vec e_s)\lambda_s^{Pt}\Theta_s\tanh \frac{d_{Pt}}{2 \lambda_s^{Pt}},
\label{Jc}
\end{equation}
where $\vec e_n$ is a unitary vector normal to the interface, $\vec e_s$ is the polarization direction of the spin current, $d_{Pt}$ is the thickness of platinum layer, $\lambda_s^{Pt}$ is the spin diffusion length in platinum, $\Theta_s$ is the spin-Hall-angle of platinum and $\mathcal{J}_s(0_+)$ is the magnitude of spin current density generated in the ferromagnet, penetrating into the heavy metal, and thus responsible for the charge current oscillations at the platinum side of the interface. At this stage, it is important to recall that the magnitude of the $\mathcal{J}_s$ propagating in platinum can be smaller than the one generated in the ferromagnet. The discontinuity of spin current between the two sides of the active interface is generally ascribed to the spin-decoherence induced by local spin-orbit fields (also known as spin-memory-loss) \cite{Jaffres} and has been recently proved \textit{via} refined spin-orbit torque experiments \cite{Buhrman2}.

\begin{figure}
\includegraphics[width=\columnwidth]{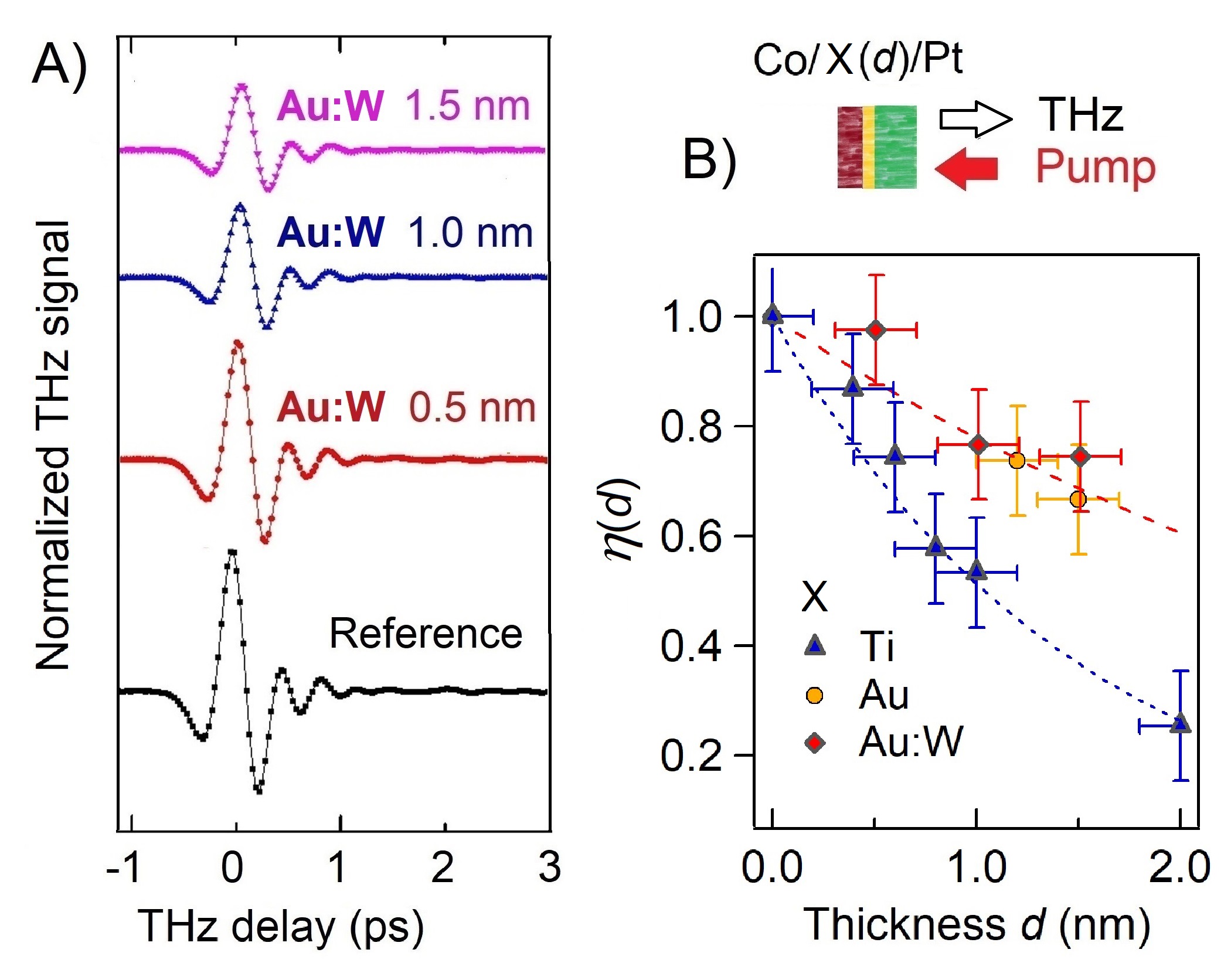}
\caption{A) THz waveforms emitted by a set of different trilayers Co/Au$_{0.85}$W$_{0.15}$($d$)/Pt. B) Spin current generated in Co/Ti($d$)/Pt, Co/Au($d$)/Pt and Co/Au$_{0.85}$W$_{0.15}$($d$)/Pt trilayers of different thickness. The Co and Pt layer have fixed thickness of 2 nm and 5 nm, respectively. The thickness $d$ of the X=Ti,Au,Au$_{0.85}$W$_{0.15}$ layer is instead varied between 0 nm and 2 nm. The parameter $\eta(d)$ has been extracted from the THz signal via Eqn. [\ref{eta}] and can be considered as a normalized spin current density in the platinum layer.}
\label{THz}
\end{figure}

Emission over an ultrabroad spectral range \cite{Seifert1,Seifert3} and theoretical modeling \cite{Schneider,Battiato,Seifert3} show that $\mathcal{J}_s$ evolves on a timescale comparable to the energy and momentum relaxation of hot electrons. We make use of the magneto-circuit formalism to write to the spin current in terms of spin conductance parameters \cite{Brataas}. The ultrafast generation of a spin accumulation on the ferromagnetic side leads to a longitudinal component whereas the spin accumulation on the Pt side induces a transverse component \cite{Seifert3}. The resulting expression reads:
\begin{equation}
\mathcal{\vec J}_s=\frac{\hbar}{4\pi} \left ( \frac{g^{\uparrow}+g^{\downarrow}}{2} \langle \frac{\partial_t M}{M}\hat{\mathbf{m}}\rangle + g^{\uparrow \downarrow}\langle \hat{\mathbf{m}}\times \partial_t \hat{\mathbf{m}}\rangle \right ),
\label{Js}
\end{equation}
where $M$ is the magnetization magnitude in the very proximity of the interface, $\hat{\mathbf{m}}$ is the local magnetization direction, $g^{\uparrow}$ ($g^{\downarrow}$) is the spin conductance parallel (antiparallel) to the magnetization and $g^{\uparrow \downarrow}$ is the spin mixing conductance. The longitudinal component is proportional to $(g^{\uparrow}+g^{\downarrow})/2$ and to the relative demagnetization $\partial_t M/M$. This term is driven by the quasi-ballistic transport of highly excited electrons from the Cobalt to Platinum and it represents the dominant contribution in the case of the spin current that are generated by ultrafast laser pulse \cite{Battiato}. Accordingly, the THz emission from Co/Pt is many orders of magnitude more intense \cite{Seifert1} than the one observed from an interface where the longitudinal component is inactive \cite{Seifert3}.

Owing to the quasi-ballistic nature of the injection, the spin current arises the spin accumulation taking place on a length scale $1.4$ nm \cite{Seifert3,Seifert2,Papaioannou}. As a consequence, the strength of the emitted THz radiation scales as the energy density injected by the pump pulse \cite{Seifert1,Seifert2,Papaioannou}
\begin{equation}
|\frac{\partial_t M}{M}|\propto\frac{A_B F_I}{d+d_{Pt}+d_{Co}},
\label{Spin}
\end{equation}
where $A_B$ is the absorbed fraction of pump pulse in the multilayer, $F_I$ is the incident fluence of the pump pulse, $d_{Pt}=5$ nm is the thickness of platinum layer, $d_{Co}=2$ nm is the thickness of cobalt layer, $d$ is the thickness of the X = Ti, Au or Au$_{0.85}$W$_{0.15}$ layer. To investigate this, we have prepared Co/X($d$)/Pt trilayers on glass and highly resistive Si(111) substrates by sputtering deposition at room temperature with standard experimental conditions. The Au$_{0.85}$W$_{0.15}$ material has been obtained via the evaporation of a rod containing 85\% of gold and 15\% of tungsten. The thickness $d$ of the interlayer is typically varied between 0 and 2 nm. Within this range of $d$, the $A_B$ coefficient can be considered constant \cite{Seifert1,Seifert2,Papaioannou}. Morever the incident laser fluence $F_I$ has been kept fixed and stable.

The THz TDS system is placed in a reflection geometry where the generated THz pulses are collected from the same surface of the spin-emitter as the excitation (i.e. no beam passes through the substrate). The emitters are mounted with small magnetic field parallel ($\cong 10$ mT) to the spin interface. We verified that a switching of the $\vec M$ orientation reverses the direction of the emitted THz field, thereby confirming that charges currents arise from the ISHE. Fig.~\ref{THz}A) displays a set of THz traces emitted from Co/Au$_{0.85}$W$_{0.15}$($d$)/Pt multilayers with different values of the Au$_{0.85}$W$_{0.15}$ thickness $d$. The THz traces recorded for different values of $d$ hold nearly identical waveforms (see also supplementary information file \cite{Sup}). Since $E_T(t,d)\cong E_T(d)f(t)$ (and equivalently $ E_T(\omega,d)\cong E_T(d) f (\omega)$), we assume that spin fluctuations, spin mixing conductance and spin-Hall-angle have negligible frequency dependence within the bandwidth of the detected THz. As observed experimentally, the drop of THz signal as a function of $d$ is mainly due to a decreasing spin-conductance. The latter is related to the detection of the THz field via Eqs.~[\ref{Imp}-\ref{Spin}].
By solving for the spin conductance we obtain:
\begin{equation}
\eta(d)=\frac {g^{ \uparrow}(d)+g^{ \downarrow}(d)}{g^{\uparrow}(0)+g^{\downarrow}(0)}=\frac{E_T(d)}{E_T(0)}\frac{Z(0)}{Z(d)}\frac{d+d_{Pt}+d_{Co}}{d_{Pt}+d_{Co}},
\label{eta}
\end{equation}
where the impedance $Z(d)$ has been calculated by assuming the THz conductivity in thin films \cite{Seifert2,Papaioannou,Walther} $\sigma_{Co}=3\times 10^6$ S/m, $\sigma_{Pt}=4\times 10^6$ S/m, $\sigma_{Au}=4\times 10^6$ S/m, $\sigma_{Au:W}=1.2\times 10^6$ S/m and $\sigma_{Ti}=0.5\times 10^6$ S/m. Differences of these conductivities with respect to bulk values are due to strong charge scattering at the landscape of the interface and to the formation of small grains \cite{Walther}. As a matter of facts, the factor $Z(0)/Z(d)$ remains close to unity, owing to the fact that metallic interlayers with nanometric thickness have small parallel conductivity.

\section{Data analysis and discussion}

The parameter $\eta(d)$ of Eqn. [\ref{eta}] reflects the relative reduction of the spin-injection efficiency in Pt if an interlayer of thickness $d$ is grown between Co and Pt. As shown by Fig. \ref{THz}B), $\eta(d)$ follows nearly an exponential decay $\exp(-d/l_X)$, with characteristic length $l_{Au}=4$ nm for X=Au or X=Au$_{0.85}$W$_{0.15}$ and $l_{Ti}=1.5$ nm for X=Ti. As can be observed, Ti affects the spin mixing conductance much more effectively than Au or Au$_{0.85}$W$_{0.15}$ do. Recent experiments have shown that a submonolayer of Ti can indeed substantially modify the spin-transfer torque of the CoFeB/Pt \cite{Park} and Co/Pt \cite{zhu2019b} interfaces. The insertion of the chemically reactive Ti alters the spin dependent transmission/reflection probabilities that favor the transport of one spin flavor with respect to the other. Furthermore, the surface passivation by Ti atoms may modify the spin-flip scattering potential at the interface. Although the microscopic mechanisms leading to the large reduction of spin conductance is still debated, a systematic investigation of spin orbit torque with different transition metals concluded that the $d$-orbital filling has a stronger influence on charge-to-spin conversion than the atomic number \cite{Avci}. Our measurements corroborate this finding: the passivation of Co/Pt interface is more effective in the case of a transition metal with incomplete $3d$-shell like titanium than in the case of an alloy with larger atomic number but closed $5d$ shell like Au. Moreover, the larger spin-orbit interaction of W in the Au$_{0.85}$W$_{0.15}$ does not seem to make any appreciable difference with respect to pure gold. Our model in the last section of this article will further clarify this, somehow surprising, result.

\begin{figure}
\includegraphics[width=\columnwidth]{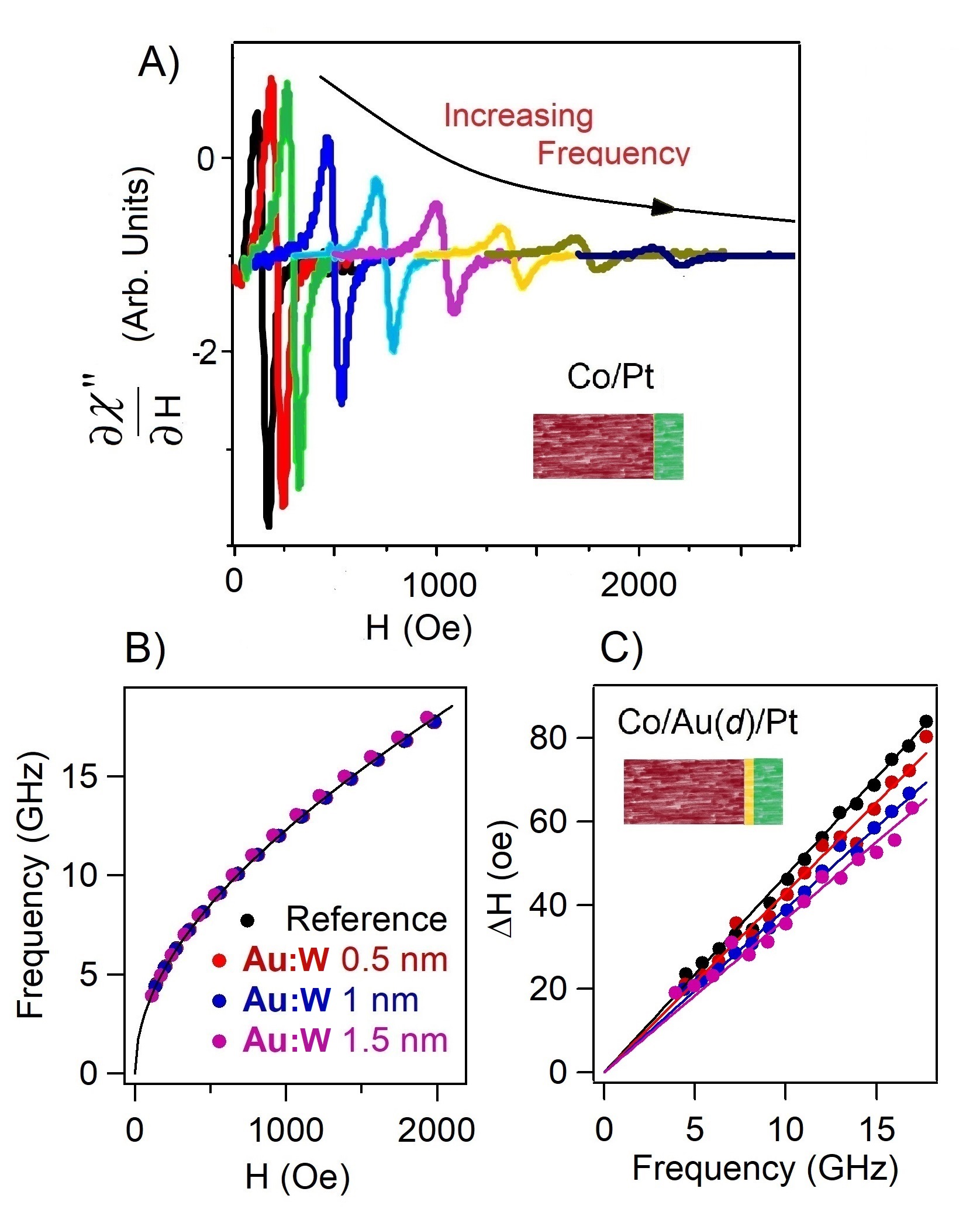}
\caption{A) Derivative of the spin susceptibility vs intensity of the static magnetic field $H$ in the reference bilayer Co/Pt. The different curves correspond to $h_{rf}$ frequencies of 4-18 GHz, with step of 2 GHz. B) Variation of resonance frequency as a function the static magnetic field $H$ in the Co/Au$_{0.85}$W$_{0.15}$($d$)/Pt trilayers. C) Full width at half maximum of the ferromagnetic resonance in Co/Au$_{0.85}$W$_{0.15}$($d$)/Pt trilayers. The Co layer has thickness of 15 nm, the Pt layer has thickness of 5 nm and the Au$_{0.85}$W$_{0.15}$ layer has thickness $d$ varying between 0 nm and 1.5 nm.}
\label{Gilbert}
\end{figure}

\begin{figure}
\includegraphics[width=0.85\columnwidth]{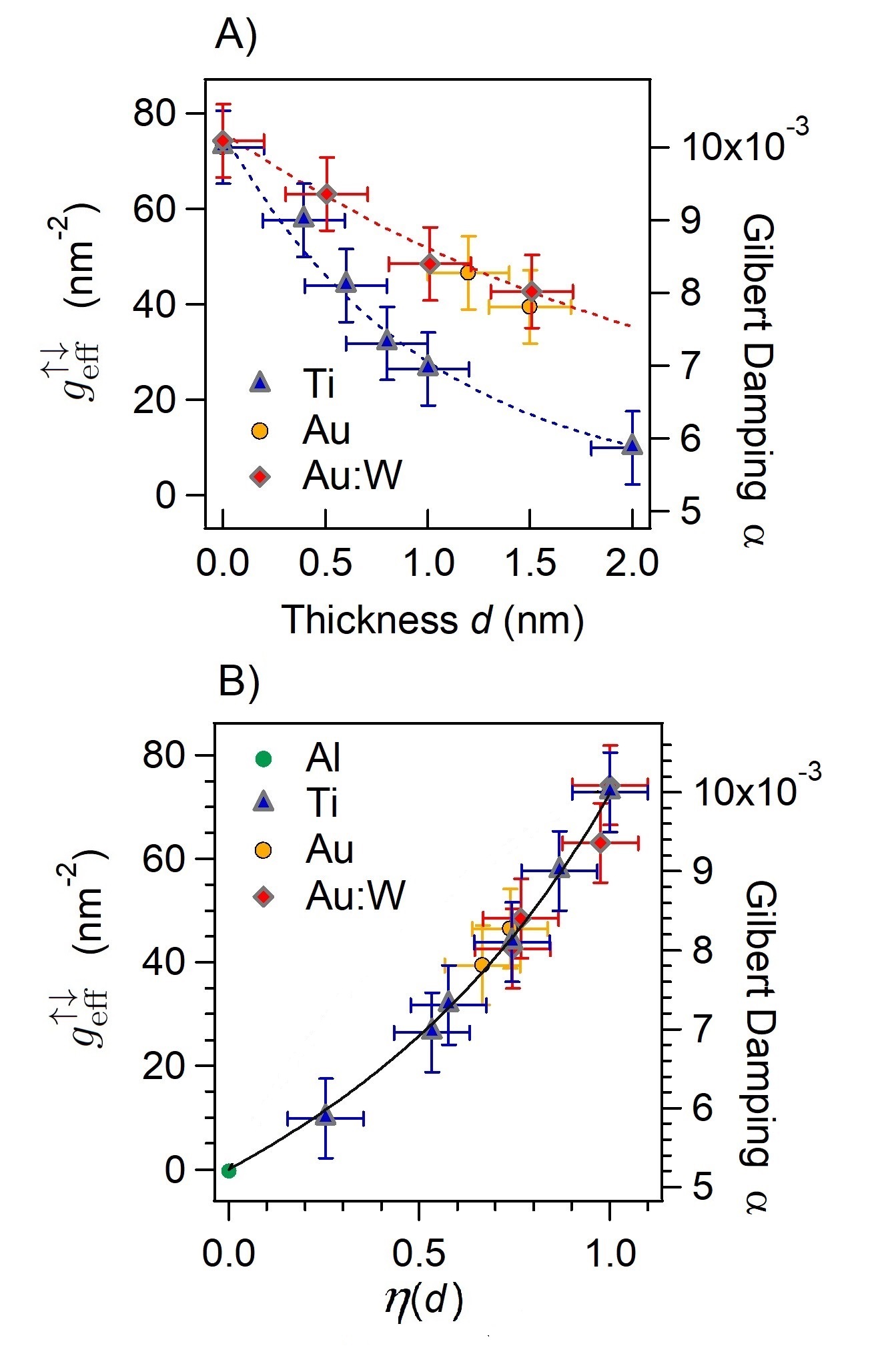}
\caption{A) Gilbert damping and spin conductance in Co/Ti($d$)/Pt, Co/Au($d$)/Pt and Co/Au$_{0.85}$W$_{0.15}$($d$)/Pt trilayers as a function of thickness $d$. B) Gilbert damping and spin conductance of the two trilayer set plot against the $\eta$ parameter extracted from the emitted THz. The green dot corresponding to vanishing THz emission is the intrinsic Gilbert damping measured on cobalt capped by 2~nm of alumina. A model that includes  the spin memory loss is calculated \textit{via} Eq.~[\ref{memory}] and superimposed (solid line) to the experimental data.}
\label{Model}
\end{figure}

Next, we discuss the effective spin mixing conductance that is measured by means of FerroMagnetic Resonance (FMR)\cite{Dang2020}. Samples made with 5~nm of Pt and thicker Co films (15nm) were deposited on highly resistive Si/SiO2(111) substrates before lithography patterning. The thicker ferromagnetic layer provides a clearer resonance spectrum compared to a 2 nm layer. Fig.~\ref{Gilbert}A) displays the differential susceptibility of the Co/Pt bilayer as a function of the external magnetic field $H$. Curves of different colors stand for increasing frequency of radiofrequency field $h_{rf}$. We show in Fig.~\ref{Gilbert}B) the resonance frequency $\omega_r$ as a function of $H$ for the multilayers Co/Au$_{0.85}$W$_{0.15}$($d$)/Pt. The FMR theory predicts:
\begin{equation}
\omega_r=\gamma\mu_0\sqrt{H(H+M)},
\label{Resonance}
\end{equation}
where $\gamma$ is the gyromagnetic ratio and $\mu_0$ vacuum permeability and $M$ is the saturation magnetization. By fitting the data with Eq.~[\ref{Resonance}], it is possible to extract the saturation magnetization $M=1500 \pm 50$ emu/cm$^3$. The damping term can be quantified by measuring the half width at half maximum $\Delta H$ of FMR linewidth. As shown by Fig. \ref{Gilbert}C) the linear regression
\begin{equation}
\Delta H=\Delta H_0 + \frac{\omega_r\alpha}{\gamma \mu_0},
\label{Width}
\end{equation}
provides the Gilbert damping $\alpha(d)$ for the Co/Au$_{0.85}$W$_{0.15}$($d$)/Pt series. Likewise, this procedure is applied to extract the Gilbert damping of Co/Ti($d$)/Pt trilayers. Moroever, the larger thickness of cobalt layer ($d_{Co}=15$ nm  in FMR experiments instead of $d_{Co}=2$ nm chosen for the THz emission experiment) minimize the extra contribution of two-magnons scattering to the $\alpha$ value. Since two-magnon scattering scales as $1/d_{Co}^2$, the associated damping term \cite{Buhrman1} should not exceed $8\times 10^{-4}$ and it has been neglected. Therefore, $\alpha$ differs from the intrinsic $\alpha_0$ only by a term arising from the injected spin current. The effective spin mixing conductance $g^{\uparrow \downarrow}_{\textrm{eff}}$ is obtained \textit{via}~\cite{Jaffres,Kelly1}:
\begin{equation}
\Delta \alpha = \alpha - \alpha_0 =\frac{g \mu_B}{4\pi M d_{Co}}g^{\uparrow \downarrow}_{\textrm{eff}},
\label{alpha}
\end{equation}
where $g$ stands for Land\'e factor of the electron and $\mu_B$ is the Bohr magnetron. The value $\alpha_0=5\times10^{-3}$ is obtained by measuring the Gilbert damping of a 15~nm cobalt capped by 2~nm of alumina.

Fig.~4A) shows $\alpha$ and $g^{\uparrow \downarrow}_{\textrm{eff}}$ for the two  trilayer series as a function of interlayer thickness $d$. Similarly to THz measurements, the drop of spin mixing conductance is faster in Co/Ti($d$)/Pt than in Co/Au$_{0.85}$W$_{0.15}$($d$)/Pt samples. This finding highlights the first important outcome of this work: an intimate connection between $(g^{ \uparrow}+g^{\downarrow})/2$ obtained by ultrafast currents in the THz spectral range, with $g^{\uparrow \downarrow}_{\textrm{eff}}$ extracted from the FMR damping linewidth. We find phenomenologically the universal relation: 
\begin{equation}
g^{\uparrow \downarrow}_{\textrm{eff}}\propto\frac{\eta(d)}{1-\xi(d)}.
\label{memory}
\end{equation}
 The solid line of Fig.~\ref{Model}B is calculated from Eq.~[\ref{memory}] with parameters $g^{\uparrow \downarrow}_{\textrm{eff}}(0)=75\textrm{nm}^{-2}$ and $\xi(d)=0.4 \eta(d)$. From their dependence on the transmission coefficient at the interface \cite{Brataas}, we evince that $g^{ \uparrow}$, $g^{\downarrow}$ and $(g^{ \uparrow}+g^{\downarrow})/2<g^{\uparrow\downarrow}\cong g^{ \uparrow}$ should scale as $\eta(d)$ upon the insertion of the interlayer. Namely, we assume that $g^{ \uparrow}(d)/g^{ \uparrow}(0)=g^{\downarrow}(d)/g^{ \downarrow}(0)=g^{\uparrow\downarrow}(d)/g^{\uparrow\downarrow}(0)=\eta(d)$. Moreover, we set $g^{\uparrow \downarrow} = (1-\xi) g^{\uparrow \downarrow}_{\textrm{eff}}$, where the parameter $\xi<1$ arises from the spin-memory-loss~\cite{Jaffres,Kelly1,Kelly2}. Due the spin scattering at the interface, the spin-current leading to THz emission in platinum is $1-\xi$ times smaller than the spin current affecting the $\vec M$ precession. The second important result of our work is that $\xi$ is proportional to the spin conductance at the interface. The more efficient the generation of spin current, the higher the spin memory loss. When expressed in terms of relative variation of spin conductance, the spin memory loss $\xi$ appears to be insensitive to the compound and thickness that has been employed to perform the passivation of the interface. We now turn on to the modeling of the spin memory loss through Rashba fields at the interface.

\section{Modeling of spin memory loss through Rashba spin-orbit interaction at the interface.}

\subsection{Electronic quantum transmission with spin-orbit interaction}

The insertion of an interlayer X at the Co/Pt interface has two mains effects: \textsl{i)} the formation of a thin potential barrier is accompanied by smaller the spin-transmission \textit{vs} Co/Pt. Indeed Co/Pt is known to build an excellent matching for the majority spin channel near the Fermi level whereas a larger chemical mismatch may take place in the case of Co/X/Pt with X=Ti, Au or Au$_{0.85}$W$_{0.15}$ and; \textsl{ii)} since the Ti or pure Au lack the open 5\textit{d} shell of Pt, the presence of an interlayer has to reduce spin orbit interaction (SOI) at the interface~\cite{Dolui}.

In the following, we consider a simplified SOI assisted quantum transmission model that has been recently implemented with success for the description of SOT~\cite{stiles2013,kim2017,borge2017,amin2018,haney2020}. This model will first highlight the role of \textsl{i)} and \textsl{ii)} in the description of our data. The interface is treated as an ideal trilayer structure Co/X/Pt with a spin current $\mathcal{J}_s$ propagating along the $\vec e_n$ direction, normal to the layers (CPP geometry). $\mathcal{J}_s$ is computed from the propagation of selected plane waves with in-plane \textit{conserved} wavevector $k_\parallel$, and normal wavector $k_z$ along $\vec e_n$. The quantum transmission is summed hereafter over the Fermi surface, as it is required within an extended Landauer treatment. We obtain the $\mathcal{J}_s(z)$ profile across the interface \textit{via} a refined model involving a Rashba-like term \cite{stiles2013,kim2017,borge2017,amin2018}. We restrict the electronic states to two electron bands with spin polarized states. The partitioned Hamiltonian in Co and Pt reads:
\begin{equation}
    \hat{\mathcal{H}}=\frac{\hat{p}^2}{2m^*}+\Delta_e \hat{\mathbf{m}}\cdot\hat{\mathbf{\sigma}}+\hat{V}
		\label{Hamiltonian}
\end{equation}
where $\hat{p}=-i\hbar \nabla_z$ is the impulsion operator, $m^*$ is the effective mass, $\hat{\mathbf{m}}$ is the magnetization direction, $\Delta_e \simeq 2$~eV is the exchange coupling for Co, and $\hat{V}=\hat{V}_{Co}=0$ represents the energy position of the bottom of the spin-averaged 3\textit{d} Co bands. Along the same lines, we set for Pt an exchange coupling $\Delta_e=0$ and $\hat{V}=\hat{V}_{Pt}\simeq -1$ eV. The potential difference $\hat{V}_{Co}-\hat{V}_{Pt}$ is representative of the workfunction offset between the two metals.

The addition of an interlayer is simulated by an interfacial potential $\hat{\mathcal{V}}_S$ that is expressed by~\cite{kim2017,amin2018,haney2020}:
\begin{equation}
    t_I\hat{\mathcal{V}}_{S} \delta(z)=t_I\left[\mathcal{V}_X  + \frac{\alpha_{R}}{\hbar}\left(\hat {\vec{p}}\times \vec e_n \right)\cdot \hat{\mathbf{\sigma}}\right]\delta(z),
\end{equation}
where $z$ is the coordinate along the direction $\vec e_n$, the function $\delta(z)$ is Dirac delta function and $t_I$ is the effective interface thickness. The operator $\hat{\mathcal{V}}_S$ is defined via: $\mathcal{V}_X$ is the average interface of an unpolarized potential barrier and $\alpha_{R}$ is the strength of Rashba interaction. We introduce the two parameters having the dimension of inverse length. The quantity $k_X=\mathcal{V}_X t_I m^*/\hbar^2$ tunes the transmission trough the barrier and $k_{\textrm{so}}=\alpha_R \overline{k}_F t_I m^*/\hbar^2$ rule the strength of the spin-orbit scattering (see also supplementary information file \cite{Sup}).

\begin{figure}
\includegraphics[width=0.9\columnwidth]{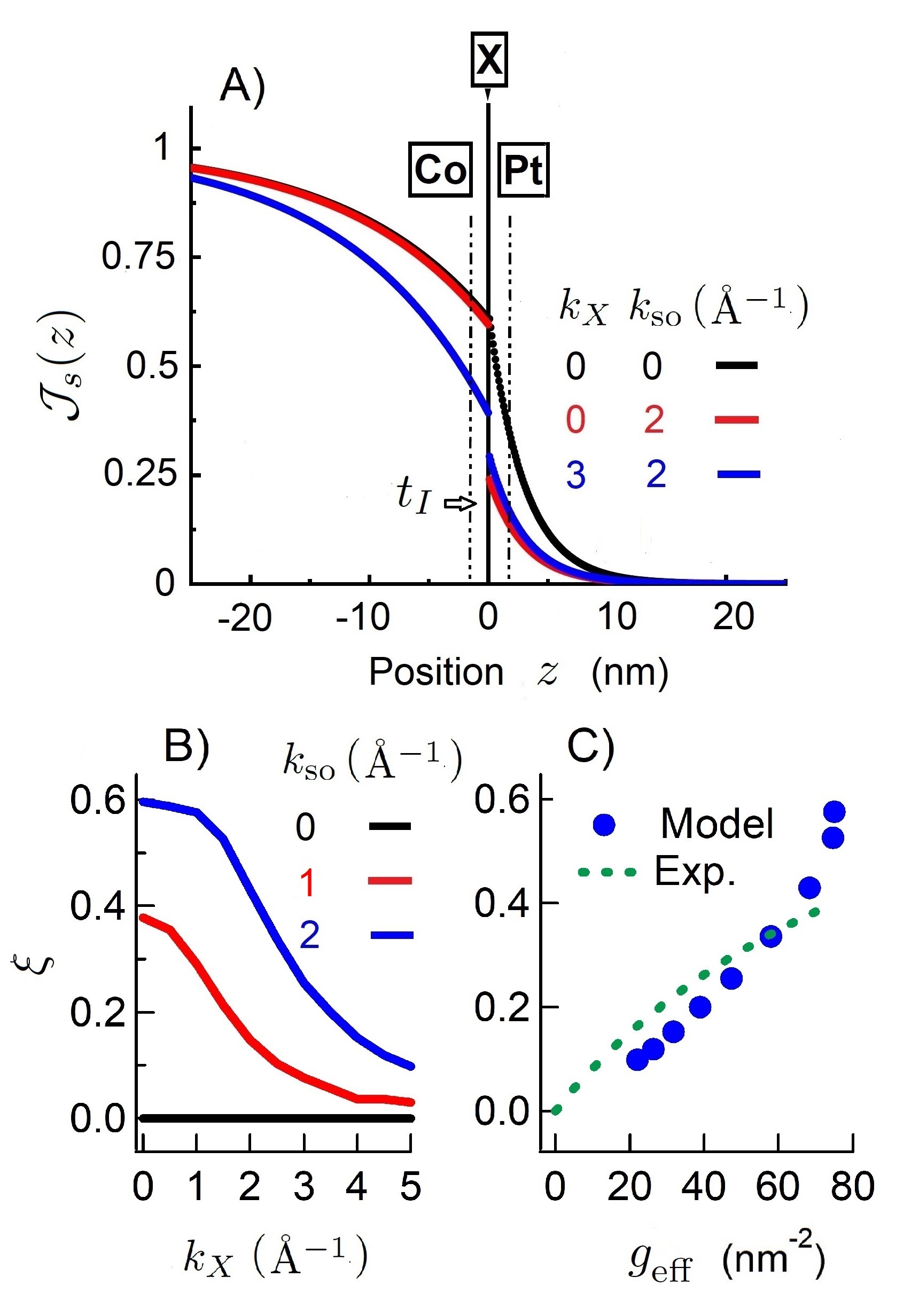}
\caption{A) Profile of spin current $\mathcal{J}_s$ in the Co/X/Pt surface at the vicinity of the Co/Pt interface for 3 different cases: no scattering potential (black curve), Rashba scattering only with $k_{\textrm{so}}=2$\AA$^{-1}$ (red curve) and Rashba scattering plus a potential barrier ($k_X=3$\AA$^{-1}$). The spin memory loss $\xi=(\mathcal{J}_s(0_-)-\mathcal{J}_s(0_+))/\mathcal{J}_s(0_-)$ is the relative discontinuity of $\mathcal{J}_s$ at the interface. B) Spin memory loss $\xi$ as a function of potential barrier $k_X$ for three different strengths of the Rashba scattering. C) Spin memory loss $\xi$ as a function of effective spin-conductance $g_{\textrm{eff}}$ obtained by varying $k_X$ and with spin orbit parameter equal to $k_{\textrm{so}}=2$\AA$^{-1}$ (blue circles). As a term of comparison we also show the relation extracted from the experimental data (green dashed line).}
\label{SpinCurrent}
\end{figure}

\subsection{Results of the model}

Our model provides the profile of a normalized spin-polarized current originating from Co (where it is normalized to unity) and propagating through a Co/X/Pt trilayer. Figure~\ref{SpinCurrent}A) depicts three specific cases, corresponding to: no interfacial potentials ($k_X=0$ and $k_{\textrm{so}}=0$), a pure Rashba interaction ($k_X=0$ and $k_{\textrm{so}}=2$\AA$^{-1}$) and, both a potential barrier and a Rashba interaction ($k_X=3$\AA$^{-1}$ and $k_{\textrm{so}}=2$\AA$^{-1}$). The spin current is always maximal in the bulk of Co, while it goes towards zero when penetrating in the non-magnetic Pt layer and moving away from the interface. In the absence of the scattering potential $\hat{\mathcal{V}}_{S}$ (black curve in Fig.~\ref{SpinCurrent}A)) the $\mathcal{J}_s(0)$ value at the Co/Pt interface results from an equilibrium condition between bulk spin-flip rates in the two regions. The spin-current is continuous everywhere (no spin-orbit scattering) and its value $\mathcal{J}_s(0)\approx 0.6$ coincides with the prediction of a pure diffusive spin-model. This agreement corroborates the validity of our quantum transmission model in the absence of any $\hat{\mathcal{V}}_{S}$ scattering.

Adding a Rashba interaction $k_{\textrm{so}}=2$\AA$^{-1}$ (red curve in Fig.~\ref{SpinCurrent}A)) leads to the spin-memory loss. Indeed the Rashba fields are not collinear to the incoming spin and induce a local spin-precession. Only a fraction of spin current coming from the Co reservoir is injected into the Pt layer so that $\mathcal{J}_s(z)$ displays a sizable discontinuity at the interface~\cite{Kelly2}. In order to quantify this effect,  we introduce the memory loss parameter $\xi=(\mathcal{J}_s(0_-)-\mathcal{J}_s(0_+))/\mathcal{J}_s(0_-)$, where $0_-$ and $0_+$ are the limiting values reached by approaching the interface from the Co and Pt side, respectively. From the chosen parameters we extract $\xi=0.6$, which is only 50\% higher than our experimental value and in agreement with previous FRM estimates \cite{Jaffres}.

The presence of an additional unpolarized scattering potential with $k_{X}=3$\AA$^{-1}$ (Blue curve in Fig.~\ref{SpinCurrent}A) has two main effects. On one hand, the larger backflow of $\mathcal{J}_s$ in the Co layer leads to a smaller ejection of spin-current from the ferromagnet. On the other hand, an unchanged strength of the Rashba field results in a smaller jump of the $\mathcal{J}_s$ current at the interface. As shown in Fig. \ref{SpinCurrent}B), the monotonic reduction of spin memory loss as a function of $k_{X}$ takes place for two representatives values of the inverse spin length $k_{\textrm{so}}$.

We extract the effective spin conductance from the rescaled ratio between the spin current $\mathcal{J}_s(0_-)$ obtained in the presence of an interlayer (i.e. for $k_{X}>0$) and the $\mathcal{J}_s(0_-)$ obtained for the bare Co/Pt interface (i.e. for $k_{X}=0$). Figure \ref{SpinCurrent}B) shows the calculated $\xi$ \textit{vs.} $g_{\textrm{eff}}$ when the potential barrier $k_X$ is increased linearly to 5\AA$^{-1}$ while the value $k_{\textrm{so}}$ is kept fixed to 2\AA$^{-1}$. Note that the spin memory loss display the same trend of the curve that is extracted by combining FMR-spin-pumping and THz methods (green dashed line). This shows that an interposition of Ti, Au or Au$_{0.85}$W$_{0.15}$ introduces a chemical barrier at the interface. The enhanced backward diffusion of electrons has the effect of decreasing both the spin mixing conductance and the spin memory loss. This effect takes place even if the spin dependent scattering $k_{\textrm{so}}$ remains equal to the pristine value.

\section{Conclusions and Acknowledgments.}

In conclusion, we report that the spin-conductance can be extracted from broadband THz spectroscopy. The investigation of Co/X($d$)/Pt trilayers with X=Ti, Au and Au$_{0.85}$W$_{0.15}$ show that in all cases, an interlayer reduces the spin-to-charge conversion. THz experiments have been bench-marked with the effective spin-mixing conductance extracted by FerroMagnetic Resonance measurements. A model including spin memory loss show that the relative drop of spin current at the interface is proportional to the spin conductance and attains $\xi=0.4$ at the Co/Pt interface. The simulations indicate that modified spin transmission probabilities at the interface can explain this correlation. Our findings are very general and show that a combination of THz emission with FMR spectroscopy can bring accurate characterizations and provide new insights into spintronic multilayers.

\vspace{0.1in}

We acknowledge E. Jacquet for his contribution in the thin film growth and M. Cosset-Cheneau for his help in the FMR experiments. We are very thankful to Tobias Kampfrath and Marco Battiato for the enlightening discussions on the interpretation of THz emission mechanism. Synchrotron Soleil hosts a THz setup where some transmission measurements have been done. Financial support has been provided by the DGA project ITEHR (No. 2018600074) as well as ANR Project TOPRISE No. ANR-16-CE24-0017. We acknowledge the Horizon2020 Framework Programme of the European Commission under FET-Proactive Grant agreement No. 824123 (SKYTOP). This project  has received funding from the H2020 research and innovation programme s-Nebula under grant agreement No.0863155.

\end{document}